\documentclass[12pt]{iopart}

\usepackage{graphicx}
\usepackage{bm}

\usepackage{graphicx}

\begin{document}
\def \wa{\'{a}}   
\def \we{\'{e}}   
\def \wi{\'{\i}}   
\def \wo{\'{o}}   
\def \wu{\'{u}}   
\def \wn{\~{n}}

\def \beq{\begin{equation}}   
\def \eeq{\end{equation}}   
\def \beqar{\begin{eqnarray}}   
\def \eeqar{\end{eqnarray}}

\title{Work-Energy theorem in rotational reference frames}

\author{D.M.Fern\wa ndez}%
 \address{Instituto de Ciencias, Universidad de General Sarmiento, 
J.M.Gutierrez 1150, (1613) Los Polvorines, Buenos Aires, Argentina.}
\ead{diegomarianof@yahoo.com.ar}%

\author{M.F.Carusela}
\address{Consejo Nacional de Invesigaciones Cient\wi ficas y T\we cnicas, CONICET\\
Instituto de Ciencias, Universidad de General Sarmiento, 
J.M.Gutierrez 1150, (1613) Los Polvorines, Buenos Aires, Argentina.}
\ead{flor@ungs.edu.ar} 


\author{C.D.El Hasi}
\address{Instituto de Ciencias, Universidad de General Sarmiento, 
J.M.Gutierrez 1150, (1613) Los Polvorines, Buenos Aires, Argentina.}
\ead{claudio@ungs.edu.ar}

\bigskip

\begin{abstract}
In standard textbooks of college physics, the Work Energy Theorem is usually presented for inertial frames of references and it is clear that energy is conserved when there is not net work of interaction forces. But what happens when energy and work are calculated in a non inertial frame of reference? This important issue is frequently avoided. Recently an extension of the theorem was derived for reference systems in traslational motion. Here we address the theorem for two observers in relative rotation showing explicitly the differences for them. We illustrate the problem with practical examples.

\end{abstract}


\maketitle


\section{Introduction}
\label{sec:intro}

The concept of energy is one of the most fundamentals and extended in all branches in Physics. Particularly the Work Energy Theorem (WET) plays an important role in Mechanics, eventhough the idea that the work made by the net force applied on a particle give rise to the change of its kinetics energy is not a general principle, but is based on the definition of work and the Newton's second law. 

On the other hand a very relevant issue is the comparison of the Newton's Law made by two observers in relative motion, this leads to the Galilean's principle of relativity. The invariance of Newton's Law is fully analized in most textbooks of elementary physics, but usually the work energy balance is stated in a given reference frame without studying its galilean invariance.

Recently the problem of the validity of WET under change of reference systems, both inertial and non inertial ones, has been addressed \cite{camarca1}, \cite{diaz}. It is not always straightforward to extend the physical laws from one observer to another in relative motion, even worse it is difficult to make further connections between physical concepts. 
An extension to non uniform translational motion shows the difference of kinetic energy ΔK and mechanical work W between inertial and non inertial observers. An expression for the work done for inertial forces $W_{inertial}$ was given in \cite{diaz} for the case of a system of particles. Under supposition of uniform translational motion (Galilean invariance) it has been stated the deep relationship among the WET and the impulse Theorem \cite{camarca1}. Notwithstanding the wide variety of systems in rotational motion, such as earth, as far as we know there is no extension of these results for changes of coordinates to rotational (non inertial) reference systems. Generally results drawn in translational motion are not straightforwardly applied to rotating systems as can be seen in most textbooks \cite{Serway},\cite{Feyman}, where rotational dynamics is only treated after a carefull study of translational mechanics.

The presentation of the concept of energy and its conservation in introductory physics courses is a major problem eventhough several approches have been given \cite{Arons} to address it both in classics and relativistic theories\cite{camarca2}.
In the context we are dealing with, energy will be kinetic and the action of the interaction forces will be taken into account as the total work they do on the system. 
It is well known that is possible to give several definitions of work as can be pointed in \cite{Sherwood}, \cite{Arons}. 
As we are dealing with a system of several particles it is worthly to note that the difference betweeen center of mass work and particle work, is essentially related to changes in the relative positions of the particles. This internal energy will be included in a general expression.

In the present article we show the extension of WET to non inertial rotating systems. As far as we know the conection between the content of energy of two system in relative rotation has been analyzed only in the context of theoretical mechanics(\cite{landau}). Here we derive similar expression using a more intuitive approach extending the results to a system of particles and given some illustrative examples.

 In section II we present explicitly the work done by inertial forces and discuss the different interpretations given for both inertial and non inertial observers. Section III contains several examples to illustrate our approach. In section IV we present the conclusions of our work.

\section{Formulation of the problem}

Let us suppose two reference frames one at rest (or inertial) $\Sigma$ and the other rotational (or non inertial) $\Sigma'$ sppining with an angular velocity $\omega$ relative to $\Sigma$. 

A particle of mass {\it m} will have position ${\bf r}$ in $\Sigma$ and ${\bf r'}$ in $\Sigma'$. As the origin of coordinates coincides the position of the particle will be the same in both systems, but they will be expressed in different coordinates: ${\bf r'} = \bf{M} {\bf r}$ where ${\bf M}$ represents the rotation matrix (see Appendix).

Let us suppose that there is a net force ${\bf F}$ acting on the particle, then from the point of view of an observer in $\Sigma$, the WET is stated as:
$m {\bf v}.{\bf dv} = {\bf F}.{\bf dr}$ or $dK = dW$

In general when one looks at these kinds of problems in a course of mechanics is often easier to study the energy from the rotating system. We propose here an approach that made light on the origin of the  difference of energy measured by two observers in relative rotation in connection with the work of interaction and inertial forces, discussion that is not usually addressed in most textbooks.

The observer in $\Sigma'$ sees the same force, but he has to include the so called inertial forces (Coriolis and centrifugal forces) to preserve the Newton's second law:

\beq
\label{3 bis}
m \frac{{\bf dv'}}{dt} = {\bf F}-m [{  \frac{\bf d\omega}{dt}} \times {\bf r'} +2 {\bf  \omega} \times {\bf v'}+{\bf  \omega} \times  {\bf  \omega} \times {\bf r'} ]
\eeq

.

Applying the usual definition of work in the rotating frame the WET still holds and is stated as: \cite{camarca1,diaz1}.

\beq
\label{4}
m {\bf dv'}. {\bf v'} =  [{\bf F}-{ m \frac{\bf d\omega}{dt}} \times {\bf r'}-2m {\bf  \omega} \times {\bf v'} -m{\bf  \omega} \times ({\bf  \omega} \times {\bf r'}) ] . {\bf dr'}
\eeq
in abbreviated form:
$dK' = dW'$

It is possible to relate the quantities measured in both systems in order to see that the work measured in $\Sigma'$ equals the work do by interaction forces (as measured in $\Sigma$) plus a term due to inertial forces:
\beq
\label{5}
dW' = dW + dW_{rot}.
\eeq
While the same happens for the Kinetic Energy:
\beq
\label{6}
dK' = dK + dW_{rot} 
\eeq
where the rotational work is expressed by (see Appendix):
\beq
\label{centr}
dW_{rot}= -{\bf d \theta}.( {\bf r} \times {\bf F})- m ( {\bf  \omega} \times  {\bf  \omega} \times {\bf r}).{\bf dr}-
m {\bf d\omega}.({\bf r} \times \frac{{\bf dr}}{dt})
\eeq
We can see that the first term is due to the work done by the momentum of the applied forces, the second one corresponds to the work done by the centrifugal force as a function of the fixed system's variables, meanwhile the third one is related to the variation of the angular velocity.

If we are dealing with a system of $N$ particles we can generalize this expression. After lenghtly algebraic calculations and considering several clues given in \cite{goldstein} we obtain:

\beq 
\label{centr1}
dW_{rot}= -{\bf d \theta}.\sum_{i=1}^{N} ({\bf r_{i}} \times {\bf F_{i}})-\frac{{\bf  \omega} . d{\bf I} . {\bf  \omega}}{2}   - {\bf d\omega}.{\bf  L^{ext}}
\eeq
where ${\bf F_{i}}$ is the force acting on each particle, ${\bf  L^{ext}}$ is the external angular momentum of the system and $d{\bf I}$ is the differential of the tensor of inertia \cite{goldstein}.

\section{Examples}

\begin{figure}
\vspace{1cm}
\begin{center}
\includegraphics[scale=1., width=5cm, height=5cm,angle=0]{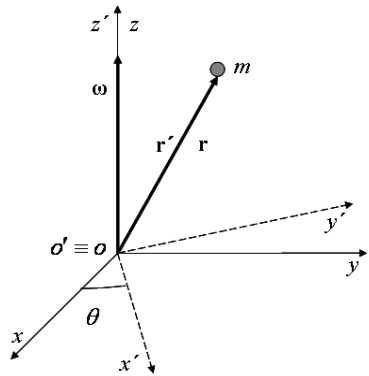}
\caption{\small{ Two references systems in relative rotational motion.}}
\label{figu1}
\end{center}
\end{figure}

In the following we present some examples showing explicitly how the different inertial terms affect the expression of the energy as seen by the two observers in relative rotation. In order to have an insight of the concepts involved but without a mess of calculations, we will considered a rotating system with constant angular velocity ${\bf \omega} = \omega \hat{z}$, we assume that the origin of both systems coincides, so that the axes {\bf z} and {\bf z'} are parallel all the time. (See Fig.\ref{figu1}).  A typical problem of this kind is the motion of a particle with respect to a system of reference fixed to Earth (non inertial frame), compared with the description from a system whose axes point to fixed stars, a better approach to an ideal inertial system \cite{goldstein}. Another interesting example would be related to the description of physical laws made by an observer standing on a carousel.

In the following examples we will mainly use eqs. (\ref{centr}) and (\ref{centr1}), where the last term is zero under the suposition that the angular velocity is constant.

\subsection{Example 1}

As a very simple example let us consider a particle of mass {\it m} fixed to a disk of neglegible mass rotating at constant angular speed. Let us suppose that the particle is held in its position due to a static friction force ${\bf F}$. From application of Newton's law in $\Sigma$ we see that ${\bf F}$ acts in radial direction pointing towards the center of the disk (the origin of $\Sigma'$). 

From Eq. \ref{centr} we observe that the first term must be zero because the force and the displacement are always perpendicular. The same argument applies to the second term. In this term it is possible to replace the product
$({\bf  \omega} \times {\bf r})$ by the linear velocity of the particle $v_{tan} \hat{\theta}$ mesured in $\Sigma$ , obtaining $-m({\bf  \omega} \times v_{tan} \hat{\theta}). {\bf dr}$. As the product ${\bf  \omega} \times v_{tan} \hat{\theta}$ and the displacement ${\bf dr}=dr \hat{\theta}$ are orthogonal, this term also vanishes. In this case $W_{rot}=0$, so the difference between $W$ and $W'$ just can be constant.
Therefore, both observers agree in the fact that the energy is conserved, even though they measured different values. One of them measure $K'=0$, meanwhile for the other $K=\frac{m v^{2}_{tan}}{2}$.

A similar result can be obtained if the particle is at rest in $\Sigma$.

\subsection{Example 2}

\begin{figure}
\vspace{1cm}
\begin{center}
\includegraphics[scale=1., width=5cm, height=5cm,angle=0]{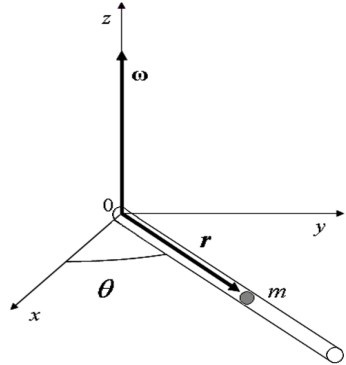}
\caption{\small{ A particle moving inside a rotating tube.}}
\label{figu4}
\end{center}
\end{figure}

Let us consider the problem of a particle moving freely in radial direction, but with a constant angular speed $\bf \omega$ such as a ball moving inside of a rotating tube with uniform circular motion like the one shown in Fig.\ref{figu4}

The interaction forces acting on the particle is only the normal force made by the tube ${\bf N}=(0,N)$ expressed in polar coordinates.
So Newton's equations seen by the observer in $\Sigma'$ can be written as:

\beq
m r \omega^{2}= m\ddot{r}
\eeq
for the radial direction, and
\beq
N-2m\dot{r} \omega=0
\eeq
for the tangential one.

Solving the radial equation leads to a solution of the form: 

\beq
\label{ro2}
r(t)=r_{o} cosh({\bf  \omega} t)
\eeq
where $r_{o}$ is the initial distance to the origin of coordinates.
The velocity can be easily calculated given:
\beq
{\bf v}=(v_{rad}(t),v_{tan}(t))=(r_{o} \omega sinh({\bf  \omega} t),r_{o} \omega cosh({\bf  \omega} t))
\eeq

Integrating Eqs.(\ref{4}) and (\ref{centr}) we get in this case:

\beq
W'= \frac{1}{2} m {\bf  \omega}^{2} r^{2} |^{r_{final}}_{r_{initial}} 
\eeq
for the work in the non inertial system, and

\beq
W_{rot}= -\frac{1}{2} m {\bf  \omega}^{2} r^{2}|^{r_{final}}_{r_{initial}} 
\eeq
for the work due to the rotation.

In principle this last expression is not zero, so the kinetic energy in $\Sigma$ is greater than that in $\Sigma'$.
The observer located in $\Sigma'$ would measured just the work done by the centrifugal force.
On the other hand the observer in $\Sigma$ would measure the same kinetic energy plus the work made to substain the system in
rotational motion.

From previous equations and Eq.\ref{5} we get:
\beq
W=  m {\bf  \omega}^{2} r^{2} |^{r_{final}}_{r_{initial}} 
\eeq

Clearly for both observers the energy is not conserved, but they do not agree on the amount of change when the particle goes from $r_{initial}$ to $r_{final}$. These results are in good agreement with the ones shown in ref.\cite{ingaard}

\subsection{Example 3}

Let us suppose a similar device as the one used in the previous example, but now the particle is also subject to a position dependent radial force, so the net force applied is: ${\bf F}=(f(r),N)$

In this case Newton's equations as seen by the observer in $\Sigma'$ can be written as:

\beq
f(r)+m r \omega^{2}= m\ddot{r}
\eeq
for the radial direction, and
\beq
N-2m\dot{r} \omega=0
\eeq
for the tangential one.

Some examples of this situation can be: a constraint force that compels the particle to move at constant speed ${\bf v'}=v_{rad}\hat{r}$, an elastic central interaction with an elastic constant $k$. For the last case after calculations we obtain that the work measured in the rotating system is:

\beq
W'= (-\frac{k r^{2} }{2}+\frac{1}{2} m {\bf  \omega}^{2} r^{2}) |^{r_{final}}_{r_{initial}} 
\eeq

$W'$ has the contributions of the elastic potential energy and centrifugal energy (work done by the centrifugal force).  What
this equation is saying is that the variation of the kinetic energy is equal to the variation, with changed sign, of the energy potential. In other words the mechanical energy in the $\Sigma'$ system is conserved.

On the other hand integrating Eq. (\ref{centr}) we obtain that:

\beq
W_{rot}= -\frac{1}{2} m {\bf  \omega}^{2} r^{2} |^{r_{final}}_{r_{initial}} 
\eeq

This expression is the result of two contributions: the work done by the momentum of the normal taking into account that is velocity dependent and the centrifugal work.

So, from Eq.\ref{5} the work as seen by an observer in $\Sigma$ is:

\beq
W= (-\frac{k r^{2} }{2}+m {\bf  \omega}^{2} r^{2}) |^{r_{final}}_{r_{initial}} 
\eeq

Let us remember that when a central force problem is solved in classical mechanics, the mechanical energy has an effective potential given by $ (\frac{k r^{2} }{2}-\frac{1}{2} m {\bf  \omega}^{2} r^{2}) $ where the last term is related to the conservation of angular momentum (\cite{landau},\cite{goldstein}).

In this case $W$ has also two contributions: one associated to the elastic potential and another one that is twice the energy potential associated to the conservation of angular momentum so that energy is not conserved. In this way both observers $\Sigma'$ and $\Sigma$ do not agree about conservation of the mechanical energy.

\subsection{Example 4}
Let us now consider a system of two particles with different masses placed on a horizontal riel rotating without friction and interacting elastically. The forces acting on each particle are the normal and elastic forces.

In order to study the pure rotational problem we place the center of mass of the system (with zero velocity) in the center of rotation and we consider polar coordinates.
The net interacting forces applied on each particle are:
\beq
{\bf F_{1}}=(-k(r_{1}-r_{2}),N_{1})
\eeq
\beq
{\bf F_{2}}=(-k(r_{2}-r_{1}),N_{2})
\eeq

Calculating the work seen by an observer in $\Sigma'$ we get:

\beq
W'= (-\frac{k (r_{1}-r_{2})^{2} }{2}+\frac{1}{2} {\bf  \omega}^{2} (m_{1}r_{1}^{2}+m_{2}r_{2}^{2})) |^{ (r_{1f},r_{2f})}_{ (r_{1i},r_{2i})}  
\eeq
where $(r_{1f},r_{2f})$ and $ (r_{1i},r_{2i})$ represents the final/initial positions of particles 1 and 2.
Integrating Eq.\ref{centr1} we get for the rotational work:

\beq 
W_{rot}= -\frac{1}{2} m {\bf  \omega}^{2}  (m_{1}r_{1}^{2}+m_{2}r_{2}^{2}) |^{ (r_{1f},r_{2f})}_{ (r_{1i},r_{2i})} 
\eeq

This work, coming from Eq.\ref{centr1}, includes two contributions. The first term is associated to the momentum of forces and gives $- m {\bf  \omega}^{2}  (m_{1}r_{1}^{2}+m_{2}r_{2}^{2}) |^{ (r_{1f},r_{2f})}_{ (r_{1i},r_{2i})}  $. The second one is related to the variations of the spatial distribution of masses in the system (time dependent moment of inertia).
Although both terms can be written as a function of the moment of inertia, the physical origin is clearly different.

When the work in $\Sigma$ is calculated similar results of example 3 are obtained. Further considerations about conservation of energy are still valid.

\section{Conclusions}
The result that the work done by the net force acting on a particle corresponds to its change in kinetic energy is a very central concept in Physics. In most textbooks it is usually stated in a given reference system and seldom related to the change of reference systems in traslational motion(\cite{diaz},\cite{camarca1}). On the other hand, the fact that there are several problems that involve rotating systems, motivated us to extend these results to the case of relative rotational motion.

We compute explicitly the difference in kinetic energy as seen for an inertial (fixed) and a non inertial (rotational) observer in an intuitive and straightforwardly way. We make the calculations for a single particle system, extending the results to systems of several particles. As far as we know this statement has never been treated before to this extent and with this scope. The only reference that we have found about this problem was faced in a more abstract form(\cite{landau}), eventhough it is only suggested the extension of the formalism to a system of particles. Here we showed in a more general way and without invoking advanced concepts from theoretical mechanics that the work done by the centrifugal force, the non uniform rotation and by the torque are the responsibles of the disagreement about conservation of energy between two observers in relative rotation. The work done by the Coriolis force results to be zero whatever the movement of the particle is.

We illustrated these results with several typical examples that help to understand the different behaviour seen by the two observers, comparing with some known results from the literature and introducing new situations. In order the make clearer the discussion we have restricted ourselves to situations with uniform rotation.

\section{Appendix}

A particle of mass {\it m} will have position ${\bf r}$ in $\Sigma$ and ${\bf r'}$ in $\Sigma'$. As the origin of coordinates coincides the position of the particle will be the same in both systems, but they will be expressed in different coordinates: ${\bf r'} = \bf{M} {\bf r}$, being {\bf M} the matrix relating the vectors in $\Sigma$ and $\Sigma' $. 

For any displacement {\bf dr} in $\Sigma$ there will be an additional rotation ${\bf d\theta} \times {\bf r}$ in $\Sigma' $, where ${\bf d\theta}$ is the angular change between both systems during the movement. In this way we get \cite{goldstein}:
\beq
\label{1}
{\bf dr'}={\bf dr}- {\bf  d\theta} \times {\bf r}
\eeq
which corresponds to the displacement in the $\Sigma' $ system. 
Because we are considering a non relativistic approach time intervals and masses are invariant under changes of referential systems, so  \cite{goldstein}:
\beq
\label{2}
{\bf v'}={\bf v}- {\bf  \omega} \times {\bf r}
\eeq
and
\beq
\label{3}
{\bf dv'}={\bf dv}- 2 {\bf  d\theta}  \times {\bf v'}-{\bf \omega} \times ({\bf  d\theta} \times {\bf r'})-{\bf d\omega} \times {\bf r'})
\eeq

Let us suppose that there is a net force ${\bf F}$ acting on the particle, then for an observer in $\Sigma'$  he has to include the so called inertial forces (Coriolis, centrifugal and the force associated to angular acceleration) to preserve the Newton's second law:

\beq
\label{3 bis}
m \frac{{\bf dv'}}{dt} = {\bf F}-m [2 ({\bf  \omega} \times {\bf v'})+{\bf  \omega} \times  {\bf  \omega} \times {\bf r'}+\frac{{\bf  d\omega}}{dt} \times {\bf r'}]
\eeq
Applying the usual definition of work in the rotating frame the WET still holds and is stated as: \cite{camarca1,diaz1}.

\beq
\label{4}
m {\bf dv'}. {\bf v'} =  [{\bf F}-2m({\bf  \omega} \times {\bf v'})-m{\bf  \omega} \times ({\bf  \omega} \times {\bf r'})-m \frac{{\bf  d\omega}}{dt} \times {\bf r'} ] . {\bf dr'}
\eeq
in abbreviated form:
$dK' = dW'$

It is possible to relate the quantities measured in both systems in order to see that the work measured in $\Sigma'$ equals the work do by interaction forces (as measured in $\Sigma$) plus a term due to inertial forces:
\beq
\label{5bis}
dW' = dW + dW_{rot}.
\eeq

We consider the work as seen by an observer in $\Sigma'$
\beq
\label{7}
dW'= {\bf F}.{\bf dr'}-2m( {\bf \omega} \times {\bf v'} ).{\bf dr'}-m ({\bf \omega} \times {\bf \omega} \times {\bf r'}). {\bf dr'}-m (\frac{{\bf  d\omega}}{dt} \times {\bf r'}).{\bf dr'}
\eeq
As the product $({\bf  \omega} \times {\bf v'} )$ is perpendicular to the displacement ${\bf dr'}$, the dot product makes null the second term. Physically this implies that the Coriolis force does not do any work in the rotational system, because it acts in a direction always perpendicular to the displacement of the particle. It would be possible to make an analogy between Coriolis and Lorentz forces: in both cases the force is the wedge product between the velocity and an external vector (the angular velocity $\omega$ or the magnetic vector field (${\bf B}$), in such a way that the force is orthogonal to the trayectory all the time and so no work is done.

Doing some algebra we get, in $\Sigma$ variables,
\beq
\label{dW3}
dW= {\bf F}.{\bf dr}
\eeq
that represents the work in the $\Sigma$ system, and

\beq
\label{dW2}
dW_{rot}= -{\bf d \theta}.( {\bf r} \times {\bf F})- m ( {\bf  \omega} \times  {\bf  \omega} \times {\bf r}).{\bf dr}-
m {\bf d\omega}.({\bf r} \times \frac{{\bf dr}}{dt})
\eeq
are the terms are related to the rotation of the system of reference.

For the case of $N$ particle systems the rotational work is:
\beq
\label{centr11}
dW_{rot}= -{\bf d \theta}.\sum_{i=1}^{N} ({\bf r_{i}} \times {\bf F_{i}})-\frac{{\bf  \omega} . d{\bf I} . {\bf  \omega}}{2}   - {\bf d\omega}.{\bf  L^{ext}}
\eeq
where
\beq
{\bf  L^{ext}}=\sum_{i=1}^{N} ({\bf r_{i}} \times {\bf p_{i}})
\eeq

is the total angular momentum and $d{\bf I}$ is the differential of the tensor of inertia of the system.

\end{document}